\documentclass{aastex631}
\usepackage{graphicx,times}
\usepackage{natbib}
\usepackage{amssymb,amsmath}
\usepackage{color}
\submitjournal{ApJ}

\shorttitle{Observation and modeling of a solar minifilament}
\shortauthors{Teng et al.}

\graphicspath{{./}{figures/}}
\begin{document}
	
	\title{HIGH-RESOLUTION OBSERVATION AND MAGNETIC MODELING OF A SOLAR MINIFILAMENT: THE FORMATION, ERUPTION AND FAILING MECHANISMS}

	\correspondingauthor{Yingna Su}
	\email{ynsu@pmo.ac.cn}
	
	\author[0000-0002-5625-1955]{Weilin Teng}
	\affiliation{Key Laboratory of Dark Matter and Space Astronomy, Purple Mountain Observatory, Chinese Academy of Sciences \\
		Nanjing, Jiangsu 210023, People’s Republic of China}
	\affiliation{Department of Astronomy and Space Science, University of Science and Technology of China \\
		Hefei, Anhui 230026, People’s Republic of China}
	
	\author[0000-0001-9647-2149]{Yingna Su}
	\affiliation{Key Laboratory of Dark Matter and Space Astronomy, Purple Mountain Observatory, Chinese Academy of Sciences \\
		Nanjing, Jiangsu 210023, People’s Republic of China}
	\affiliation{Department of Astronomy and Space Science, University of Science and Technology of China \\
		Hefei, Anhui 230026, People’s Republic of China}

	\author[0000-0003-4618-4979]{Rui Liu}
	\affiliation{CAS Key Laboratory of Geospace Environment, Department of Geophysics and Planetary Sciences, University of Science and Technology of China\\
		Hefei, 230026, People's Republic of China}
	\affiliation{CAS Center for Excellence in Comparative Planetology, University of Science and Technology of China\\
		Hefei 230026, People's Republic of China}
	\affiliation{Mengcheng National Geophysical Observatory, University of Science and Technology of China\\
		Mengcheng 233500, People's Republic of China}
	
	\author[0000-0002-2436-0516]{Jialin Chen}
	\affiliation{School of Physics and Electronic Engineering, Hanjiang Normal University \\
		18 South Beijing Road, Shiyan, Hubei, China, 442000}
	
	\author{Yanjie Liu}
	\affiliation{Key Laboratory of Dark Matter and Space Astronomy, Purple Mountain Observatory, Chinese Academy of Sciences \\
		Nanjing, Jiangsu 210023, People’s Republic of China}
	\affiliation{Department of Astronomy and Space Science, University of Science and Technology of China \\
		Hefei, Anhui 230026, People’s Republic of China}
	
	\author[0000-0003-4787-5026]{Jun Dai}
	\affiliation{Key Laboratory of Dark Matter and Space Astronomy, Purple Mountain Observatory, Chinese Academy of Sciences \\
		Nanjing, Jiangsu 210023, People’s Republic of China}
	\affiliation{Department of Astronomy and Space Science, University of Science and Technology of China \\
		Hefei, Anhui 230026, People’s Republic of China}
	
	\author{Wenda Cao}
	\affiliation{Big Bear Solar Observatory, New Jersey Institute of Technology\\
		40386 North Shore Lane, Big Bear City, CA 92314-9672, USA}

	\author{Jinhua Shen}
	\affiliation{Xinjiang Astronomical Observatory, Chinese Academy of Sciences, 150 Science 1-Street, Urumqi, Xinjiang 830011, China}
	
	\author[0000-0002-5898-2284]{Haisheng Ji}
	\affiliation{Key Laboratory of Dark Matter and Space Astronomy, Purple Mountain Observatory, Chinese Academy of Sciences \\
		Nanjing, Jiangsu 210023, People’s Republic of China}
	\affiliation{Department of Astronomy and Space Science, University of Science and Technology of China \\
		Hefei, Anhui 230026, People’s Republic of China}

	\begin{abstract}
		
Minifilaments are widespread small-scale structures in the solar atmosphere. To better understand their formation and eruption mechanisms, we investigate the entire life of a sigmoidal minifilament located below a large quiescent filament observed by BBSO/GST on 2015 August 3. The H$\alpha$ structure initially appears as a group of arched threads, then transforms into two J-shaped arcades, and finally forms a sigmoidal shape. %Unlike large scale filaments, the high-resolution observations by GST suggest that the evolution of this sigmoidal minifilament may be driven by flux feeding due to the footpoint rotation of one J-shaped branch. 
SDO/AIA observations in 171~\AA~show that two coronal jets occur around the southern footpoint of the minifilament before the minifilament eruption. The minifilament eruption starts from the southern footpoint, then interacts with the overlying filament and fails. The aforementioned observational changes correspond to three episodes of flux cancellations observed by SDO/HMI. Unlike previous studies, the flux cancellation occurs between the polarity where southern footpoint of the minifilament is rooted in and an external polarity. We construct two magnetic field models before the eruption using the flux rope insertion method, and find an hyperbolic flux tube (HFT) above the flux cancellation site. The observation and modeling results suggest that the eruption is triggered by the external magnetic reconnection between the core field of the minifilament and the external fields due to flux cancellations. This study reveals a new triggering mechanism for minifilament eruptions and a new relationship between minifilament eruptions and coronal jets.%The minifilament eruption fails partly due to the interaction with the overlying filament, and partly because it hasn't reached the threshold height of torus instability.
		
	\end{abstract}

	\keywords{Solar filament(1495) --- Solar filament eruptions(1981) --- Solar magnetic reconnection(1504)} %--- Interdisciplinary astronomy(804)}

	\section{Introduction}           
	\label{sect:intro}

Solar filaments, or when they appear as bright structures above the solar limb, prominences, are cold and dense plasma clouds that are supported and confined by magnetic fields in the solar corona. For most solar filaments, their material is stored in the dips of coronal magnetic field, where the magnetic field is horizontal and the magnetic tension force is upward \citep{1957ZA.....43...36K,1998A&A...329.1125A,2017ApJ...835...94O}. There are two types of magnetic structures proposed to produce such dips and support filament material in the corona: magnetic flux ropes, where the magnetic field is twisted around an axis \citep[$e.g.$][]{1998A&A...329.1125A,2004ApJ...612..519V}, and sheared arcades, where the three-dimensional magnetic arcades are strongly sheared along the polarity inversion line \citep[$e.g.$][]{1994ApJ...420L..41A}. Compared to the solar corona, the temperature and density of filaments are $\sim$100 times lower and $\sim$100 times higher, respectively \citep{2014LRSP...11....1P}. Among solar filaments, there are extremely small-scale ones, which are called minifilaments. They generally have a spatial scale of $\sim$19 Mm, much smaller than large-scale ones, and a shorter life cycle and a higher incidence than their large-scale counterparts \citep{2000ApJ...530.1071W}. Minifilaments are discovered in 1980s, and with the help of high-resolution telescopes and instruments, they have received more and more attention in the past decade.

Minifilaments form and erupt all over the solar disk, causing small-scale solar activities, such as coronal jets. \cite{2010ApJ...720..757M} identified two types of coronal jets: standard jets and blow-out jets. In standard jets, the emerging field simply reconnects with the pre-existing ambient field, causing hot plasma produced by magnetic reconnection to flow along the reconnected field line, producing jets  \citep{1992PASJ...44L.173S}. However, in blow-out jets, the emerging field has a potentially eruptive core structure, usually a sheared arcade, and when the magnetic reconnection between overlying field and pre-existing ambient field occurs, the core structure erupts, similar to large-scale breakout eruptions \citep{1999ApJ...510..485A}, producing blow-out jets. In this type of jets, the erupting low-lying base arch core field can carry minifilaments, and the blow-out process corresponds to the minifilament eruption \citep{2011ApJ...738L..20H,2012ApJ...745..164S,2014ApJ...783...11A,2016SSRv..201....1R}. A study by \cite{2015Natur.523..437S} suggests that minifilament eruptions is the cause of coronal X-ray jets, and different final states of minifilament eruption results in a continuum of jet morphology, ranging from standard to blow-out ones. Jet-producing minifilament eruptions are used to explain the switchback structures detected by Parker Solar Probe \citep{2020ApJ...896L..18S}. However, the \cite{2015Natur.523..437S} scenario does not specify the triggering mechanism of the minifilament eruption.

As more observations and studies of minifilament eruptions accumulate, it has been found that many of them undergo a similar physical process as large-scale filaments when they erupt \citep{2020ApJ...902....8C}. Therefore, the hypothesis that solar eruptive events are self-similar across multiple scales has gained popularity \citep[$e.g.$][]{2010ApJ...718..981R,2010ApJ...710.1480S,2022A&A...660A..45M}. However, this hypothesis is still being examined.

Various models have been proposed to explain how filaments erupt. Some models propose that ideal magnetohydrodynamic (MHD) instability triggers the eruption, such as the kink instability model \citep{2004A&A...413L..27T,2007ApJ...668.1232F,2010A&A...516A..49T} and the torus instability model \citep{2006PhRvL..96y5002K,2014ApJ...789...46K}. %In the corona, field lines can be twisted around an axis to form a flux rope, and the twist of the magnetic field results in poloidal flux and magnetic free energy. With certain disturbances, the writhe of the axis can increase, causing the entire flux rope to rise and leading to a release of free energy. This is the kink instability model. Another model suggests that the current flowing along the axis can produce an opposite image current flowing under the photosphere, and the repulsive force between them lifts the flux rope. If the lifting force decays more slowly with increasing height than the confining force by the overlying field, the torus instability can occur and accelerate the flux rope. 
Other models that consider reconnection processes propose that the eruption can be triggered by magnetic reconnection between two sheared arcades that form a flux rope \citep[the tether-cutting model,][]{2001ApJ...552..833M}, or by the magnetic reconnection of overlying fields within multipolar configurations \citep[the magnetic breakout model,][]{1999ApJ...510..485A}. 

To better understand solar activities, whether eruptive or not, a deeper understanding of the coronal magnetic field is necessary. However, direct measurements of the 3D magnetic field in the solar corona are difficult, thus, different methods have been developed to extrapolate or reconstruct the coronal magnetic field based on magnetograms on the bottom boundary. Among them are potential field extrapolation \citep[$e.g.$][]{1969BAAS....1Q.288N}, linear force-free field extrapolation \citep[$e.g.$][]{Seehafer:1978tz}, non-linear force-free field (NLFFF) extrapolation using photospheric vector magnetograms \citep[$e.g.$][]{1990ApJ...362..698W} or reconstruction using photospheric line-of-sight (LOS) magnetograms \citep[$e.g.$, flux rope insertion method,][]{2004ApJ...612..519V}, and non force-free field extrapolation \citep[$e.g.$][]{2006GeoRL..3315106H}. Due to the $180^{\circ}$ ambiguity and significant measuring errors of the photospheric transverse field, the flux rope insertion method is particularly advantageous for modeling the coronal magnetic field in regions with weak photospheric magnetic field, since it is fed solely with more reliable photospheric LOS magnetic field.

The eruption mechanism of solar filaments is an important research topic in solar physics. The lifetimes of large-scale filaments are generally long, and their formation and eruption processes are often challenging to fully observe, especially for ground-based large telescopes. However, for minifilaments, their complete lifecycle can be entirely covered by a single observation process of high-resolution solar telescopes. Therefore, studying minifilaments contributes to a better understanding of the formation and eruption mechanisms of solar filaments. Additionally, the eruption of minifilaments can disturb large-scale structures, potentially leading to large solar eruptions with space weather impacts. Moreover, as a small-scale energy release process on the Sun, minifilament eruptions are crucial in our understanding of coronal heating mechanisms. 

In this study, we investigate the formation and eruption process of a minifilament located below a large quiescent filament on August 3, 2015, using magnetic modeling and multi-wavelength high-resolution observations. The paper is organized as follows. In \S\ref{sect:inst}, we describe the observing instruments employed in this study. Analysis of multi-wavelength observations of the minifilament eruption is presented in \S\ref{sect:obser}. In \S\ref{sect:model}, we construct magnetic models to elucidate the observed phenomena and unveil the eruption mechanism. At last, we discuss the results and draw our conclusions in \S\ref{sect:disc} and \S\ref{sect:summ}.

\section{Instruments}
\label{sect:inst}

In this study, we utilize high-resolution H$\alpha$ observations taken by the Goode Solar Telescope \citep[GST,][]{2012SPIE.8444E..03G} at the Big Bear Solar Observatory (BBSO). With a combination of a high-order adaptive optics system and the post-facto speckle image reconstruction techniques \citep{2008A&A...488..375W}, GST is able to achieve diffraction-limited imaging of the solar atmosphere. Various instruments are established for GST \citep{2010AN....331..636C}, including the Fabry-Pa\'{e}rot filter-based system, Visible Imaging Spectrometer (VIS) which offers imaging spectroscopy in the wavelength range of 5500-7000 \AA. For the event understudy, we operated VIS to scan over the Ha line from -0.4 \AA~to 0.4 \AA~with a step of 0.2 \AA. The pixel size of VIS imaging is 0".029, with a cadence of 35 s. The H$\alpha$ images are co-aligned with the corresponding AIA and HMI images by matching the large-scale quiescent filament and small-scale brightenings in H$\alpha$ and 304~\AA. Both image correlation and video stabilization techniques are used to realize self-registration of H$\alpha$ images.

For multi-wavelength co-analysis and magnetic field modeling, we adopt data from the Atmospheric Imaging Assembly \citep[AIA,][]{Lemen:2012wn}, and line-of-sight (LOS) photospheric magnetograms from the Helioseismic and Magnetic Imager \citep[HMI,][]{Schou:2012vf}, and both instruments are on board the Solar Dynamics Observatory \citep[SDO,][]{Pesnell:2012vy}. AIA observes the Sun in ten channels including seven EUV channels with 12-s cadence and three ultraviolet (UV) channels with 24-s cadence. The field of view (FOV) of AIA images is larger than 1.3 R$_\sun$ and the pixel size is 0.$^{\prime\prime}$6. HMI observes the full solar disk in 6173~\AA. The LOS and vector magnetograms with pixel size of  0.$^{\prime\prime}$5 are taken every 45 s and 720 s, respectively.

\section{Observation Result}
\label{sect:obser}
\subsection{Overview of the Event}

\begin{figure*}[hbtp!]
\begin{interactive}{animation}{AAS1.mp4}
\plotone{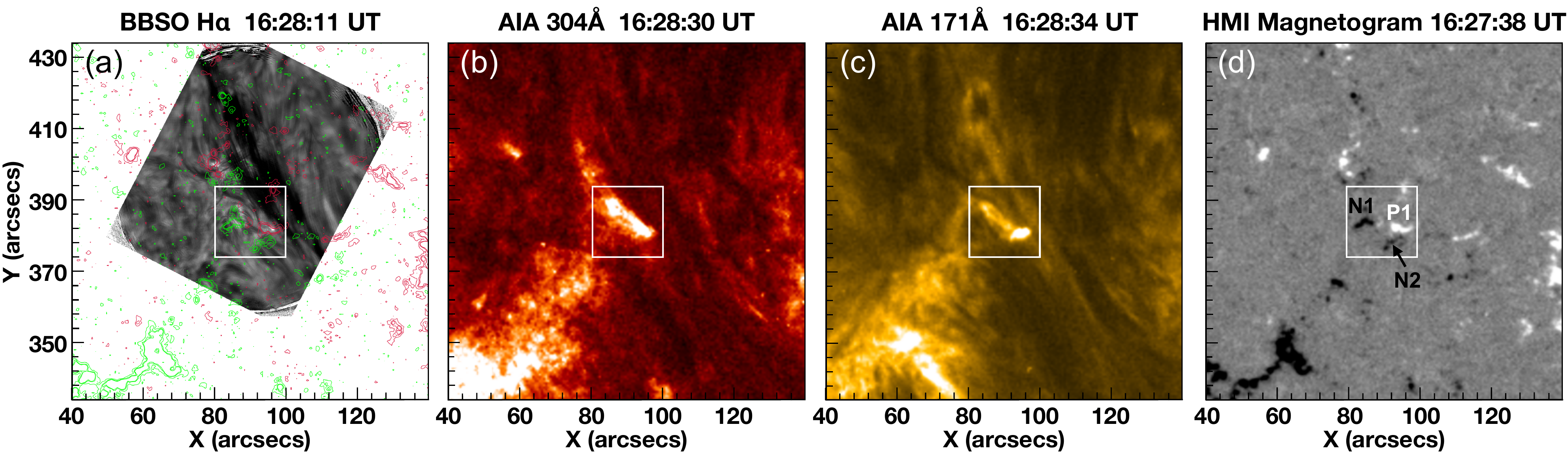}
\end{interactive}
\caption{Overview of the minifilament observed by GST (left panel) and SDO (right three panels) before eruption. (a)-(d)  Images in H$\alpha$, 304~\AA, and 171~\AA, as well as the LOS photospheric magnetogram observed around 16:28 UT on 2015 August 3. The small-scale evolving structure (the predecessor of the minifilament) is enclosed by the white boxes. The red and green contours in panel (a) refer to the positive and negative magnetic fields observed by HMI. An animation of this figure is available. It covers the time interval from 16:27 UT to 19:19 UT. The real time duration of the animation is 17 s.\\
(An animation of this figure is available.)}
\label{figov}
\end{figure*}

High-resolution observations of the minifilament was taken by BBSO/GST from 16:28 UT to 19:20 UT on 2015 August 3. Figure \ref{figov} shows the source region in multiple wavelengths and marks the position of the minifilament. At 16:28 UT, a cluster of small-scale dark arched threads is located beneath the eastern edge of the large-scale quiescent filament (Figure \ref{figov}(a)), with their two ends rooted in opposite magnetic polarities (N1 and P1 in Figure \ref{figov}(d)). There is a negative polarity (N2) next to the positive polarity in the southwest (P1), forming a pair of closely located opposite magnetic polarities. Figures \ref{figov}(b) and \ref{figov}(c) display the AIA images of the same region in 304~\AA~and 171~\AA, respectively. The EUV structure corresponding to the dark H$\alpha$ arched threads continuously brighten up, indicating occurrence of activities, which can also be clearly seen in the online animation associated with Figure \ref{figov}. Before 18:00 UT, these H$\alpha$ arched threads evolve into two J-shaped arcades and then form a sigmoidal minifilament. From 18:05 UT to 18:20 UT, the minifilament connects with an external dark structure at its southwest end and gradually erupts from south to north. After 18:14 UT, brightening and outflows near the eruption site are observed in the EUV channels. %The eruption appears to be unsuccessful, given its obvious deceleration and halt at higher altitudes, and it also disturbs the overlying large-scale filament. 

\subsection{Evolution before Eruption}
\subsubsection{Formation of the Sigmoidal Minifilament}

\begin{figure*}[hbtp!]
%\begin{interactive}{animation}{AAS2.mp4}
\plotone{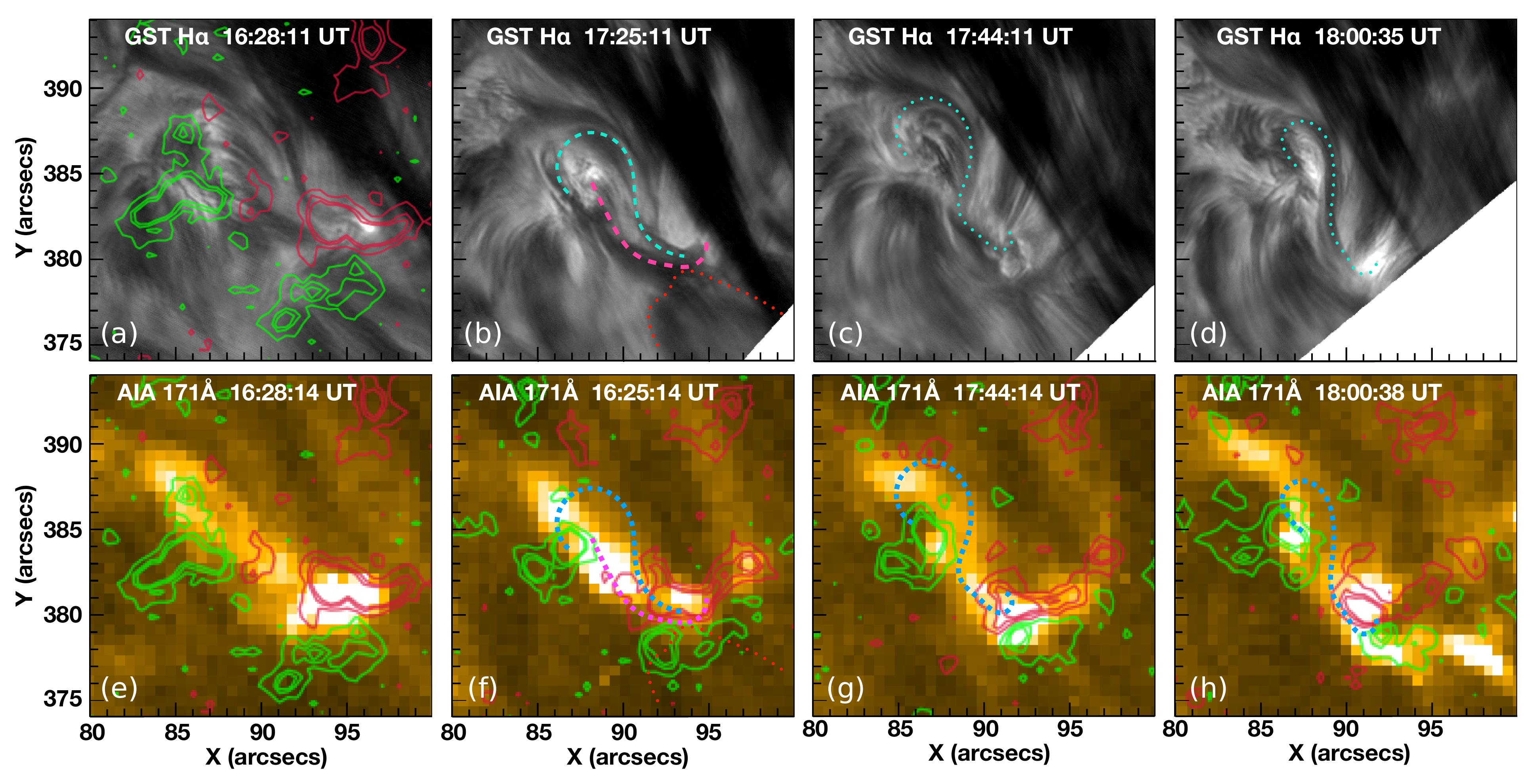}
%\end{interactive}
\caption{Formation of the minifilament observed in H$\alpha$ by GST (top row) and 171~\AA~by AIA ( bottom row). The major structures in H$\alpha$  are traced by the dashed blue and pink lines (top row), which are superposed on the corresponding 171~\AA~images (bottom row). The red and green contours in panels (a) and (e)-(h) refer to the corresponding positive and negative magnetic fields observed by HMI, with contour values of [-70, -50, -20, 20, 50, 70] Gauss.} 
%(An animation of this figure is available.)}
\label{figev}
\end{figure*}

Figure \ref{figev} depicts the miniature structure at four significant time instants in H$\alpha$ and 171~\AA. At 16:28 UT, a group of dark arched-shaped threads constitutes the small-scale H$\alpha$ structure, as illustrated in Figure \ref{figev}(a). From 16:28 to 17:25 UT, the dark threads evolve into two branches of J-shaped arcades, marked with cyan and pink dashed lines in Figure \ref{figev}(b). The southern branch connects with an external absorption structure, outlined by a red dotted line in Figure \ref{figev}(b), at its southwest end. From 17:25 UT to 17:44 UT, the southern branch of the J-shaped arcades rises with the external absorption structure and gradually vanishes, while the northern branch evolves into a sigmoidal structure (traced by the cyan dotted line in Figure \ref{figev}(c)), which is associated with the rotation of the northeast footpoint (for further discussion see Subsection \ref{subsecmf}). By 18:00 UT, the sigmoidal structure is further enhanced by filament material, and the sigmoidal minifilament is formed, as outlined by the cyan dotted line in Figure \ref{figev}(d). During the evolution of the small-scale structure, the corresponding areas in AIA 171~\AA~and other EUV passbands are dominated by bright features with time-varying brightness, indicating the ongoing magnetic activities.

\subsubsection{Bi-directional Coronal Jets}

\begin{figure*}[hbtp!]
\plotone{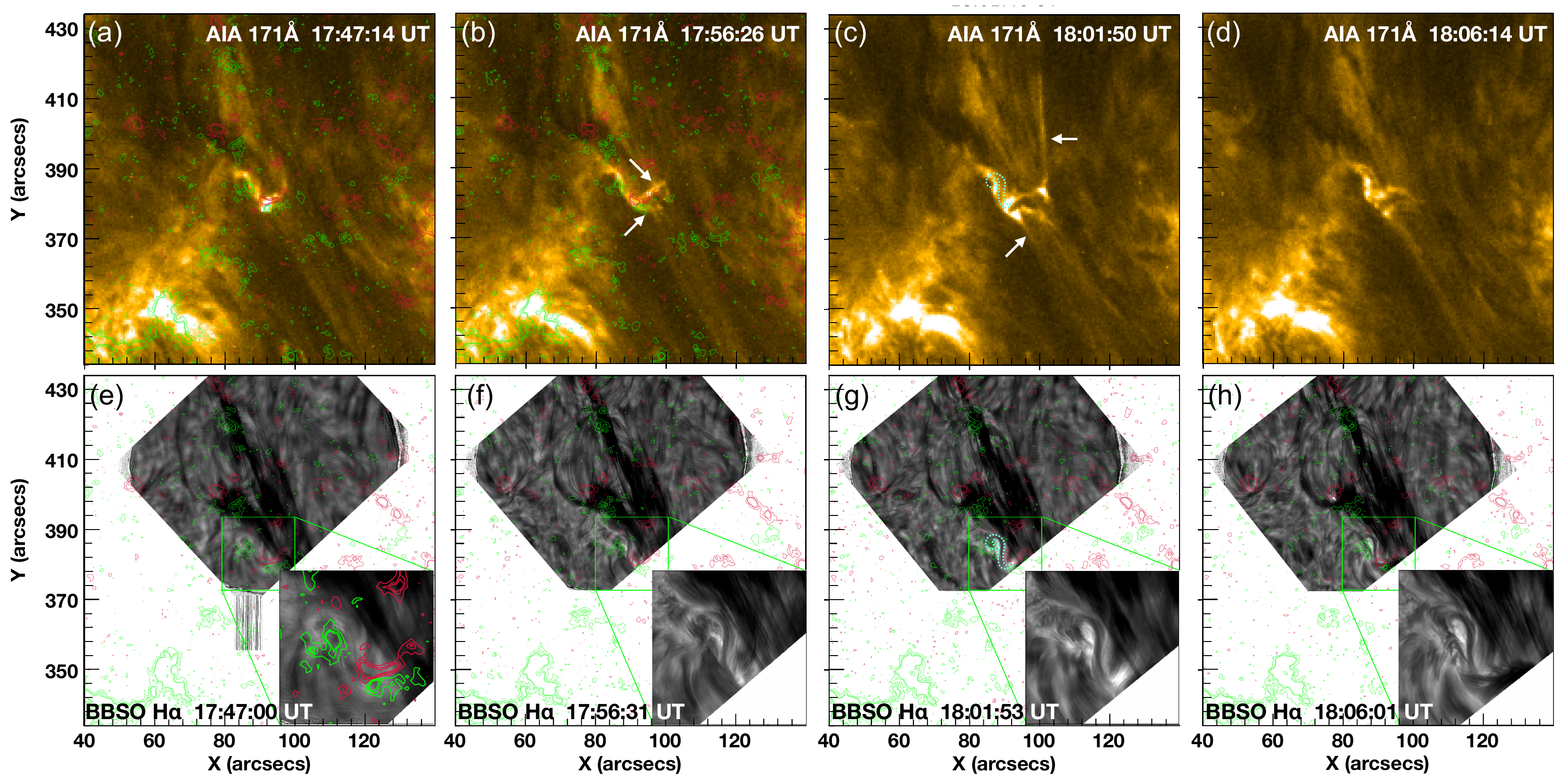}
\caption{Evolution of coronal jets. Top row: AIA 171~\AA~images with a large FOV, and the white arrows mark the bi-directional coronal jets. Bottom row: H$\alpha$ images with both large and small FOVs. The red and green contours represent the corresponding positive and negative magnetic fields observed by HMI. The sigmoidal minifilament is marked by the cyan dotted line in panel (g).}
\label{figjt}
\end{figure*}

From 17:56 UT to 18:06 UT, two EUV coronal jets are flowing out from the southwest end of the sigmoidal structure, as shown in the top row of Figure \ref{figjt}. These jets are best observed in 171~\AA, while they are also visible in other EUV passbands. At the same time, the filament material gradually fills the sigmoidal structure, as observed in H$\alpha$ shown in the bottom row of Figure \ref{figjt}, and the minifilament is forming. The coronal jets appear during the formation stage of the minifilament, prior to its eruption. The jets flow out in semi-opposite directions, generating a pair of bi-directional outflows marked with white arrows in Figures \ref{figjt}(b)-(c). The paths of the coronal jets are much longer than that of the minifilament. As the coronal jets almost disappear, the eruption of the minifilament begins.

\subsection{Minifilament Eruption}

\begin{figure*}[hbtp!]
\begin{interactive}{animation}{AAS2.mp4}
\plotone{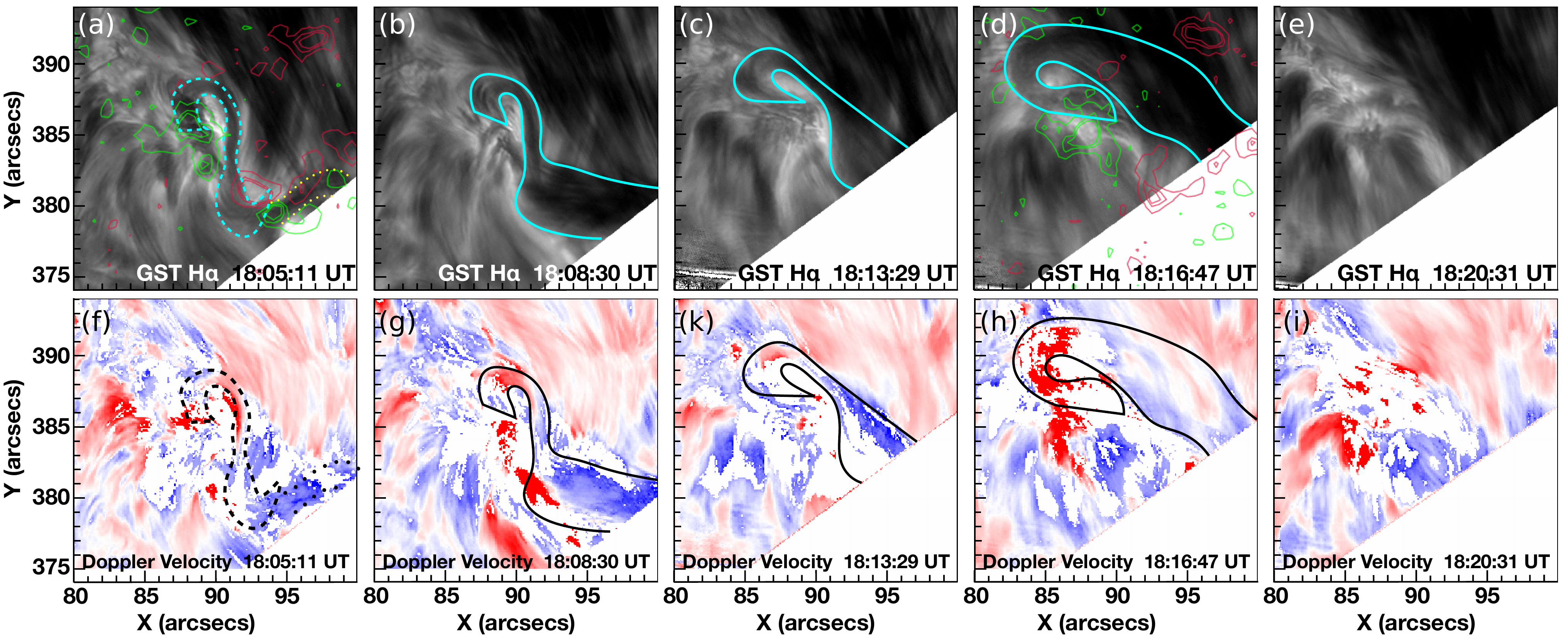}
\end{interactive}
\caption{Eruption of the minifilament. Top row: H$\alpha$ images. Bottom row: Doppler velocity maps ranging from -20 to 20 km s$^{-1}$.  Red and green contours representing the positive and negative magnetic fields taken  by HMI are superimposed on panels (a) and (d). The erupting structures are outlined with cyan and black lines in both H$\alpha$ and Doppler velocity maps. An animation of the H$\alpha$ images is available. It covers the time interval from 17:53 UT to 18:30 UT. The real time duration of the animation is 4 s.\\
(An animation of this figure is available.)}
\label{figer}
\end{figure*}

The H$\alpha$ images (top row) and corresponding Doppler velocity maps (bottom row) taken by BBSO/GST during the eruption are shown in Figure \ref{figer}. For the Doppler maps, the line center wavelength and LOS velocity of each pixel are obtained using Gaussian fitting. However, due to the limited number of wavelengths in the observation (only 5), the fitting error is relatively high, leading to a high standard deviation of the derived Doppler velocity. Therefore, we only show the region where the standard deviation is smaller than 18 km s$^{-1}$, and the rest is colored white. In Figure \ref{figer}, the main body of the minifilament is marked by cyan dashed lines, an external dark structure is traced by yellow dotted lines, and the erupting structure is shown in cyan solid lines.

From the H$\alpha$ observations we can see that the filament eruption begins at 18:05 UT, when an external dark structure quickly appears and connects to the southwest end of the minifilament. Then the southern part of the filament rises up firstly, together with the dark structure that is connected to. After that, the northern part also begins to erupt, leading to the eruption of the entire filament that ends up in an arcade halting at a much higher altitude. The whole eruption process lasts about 15 minutes. The corresponding Doppler maps show the blueshifted area extends from the southern end to almost the entire minifilament, convincingly indicating a gradual eruption from south to north. Then the decrease in absolute Doppler velocity and the appearance of red shift at the filament end show the frustration of filament eruption associated with the mass draining towards the solar surface. Thus this eruption is observed to be a failed one.

\subsection{Evolution of the Photospheric Magnetic Field}
\label{subsecmf}

\begin{figure*}[hbtp!]
\plotone{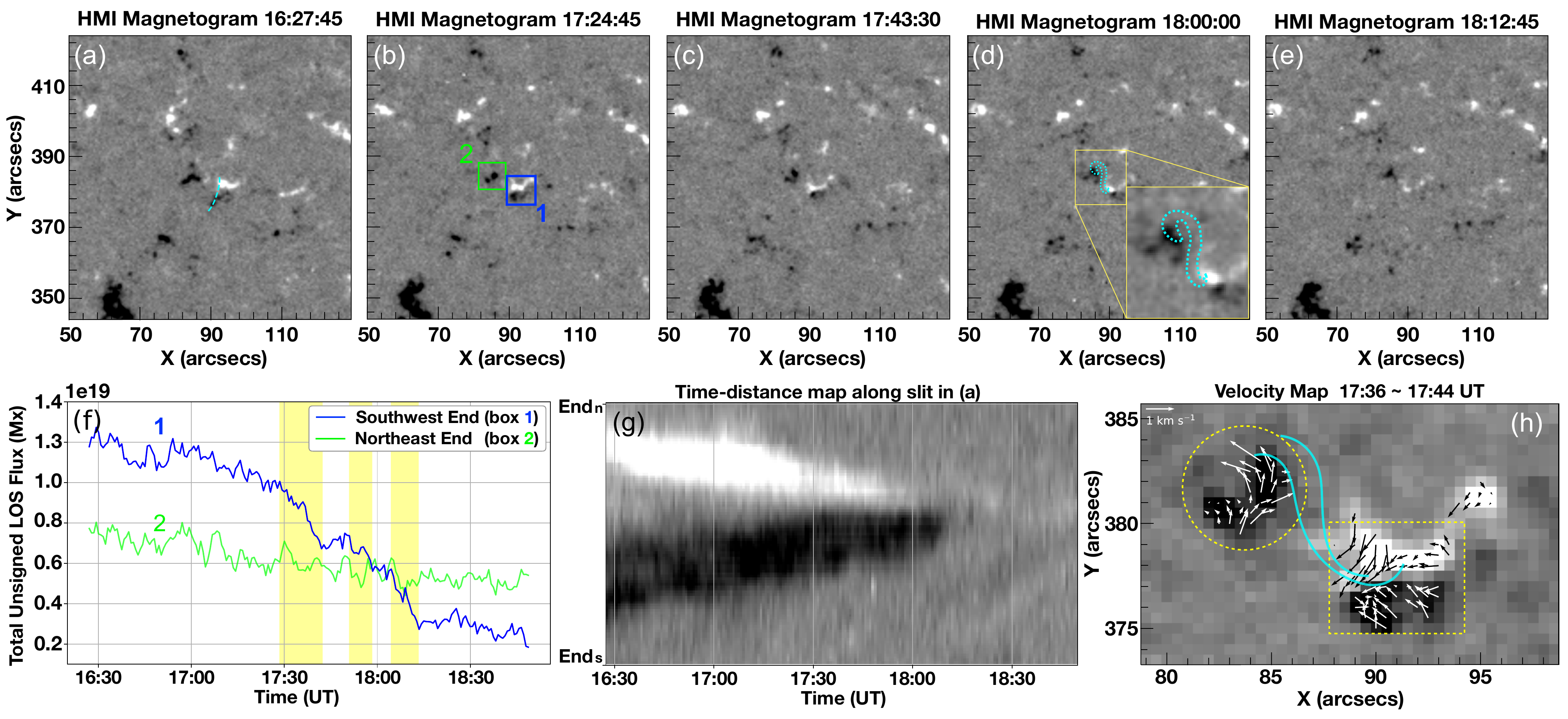}
\caption{Magnetic field evolution before and during the eruption. (a)--(e): LOS magnetograms taken by HMI at five different times. The minifilament is marked by the cyan dotted lines in panel (d). (f): Temporal evolution of the unsigned LOS magnetic flux at the southwest and northeast footpoints (enclosed by boxes 1 and 2 in panel (b)) of the minifilament. Yellow shading marks three episodes of flux cancellation in box 1. (g): Time-distance map along the cyan dashed slit in (a).  (h): Transverse photospheric flows superposed on the HMI magnetograms before the formation of the sigmoidal structure. The white and black arrows refer to the velocity vectors. The yellow circle and box encircle the regions with rotational motion and converging motion, respectively.  The H$\alpha$ structures are traced by the cyan lines.}%The red and green contours refer to positive and negative helicity flux density, respectively, with values ranging from $-10^{20}~{\rm to}~10^{20}~{\rm Mx^{2}~hr^{-1}~cm^{-2}}$. 
\label{figmg}
\end{figure*}

During the formation and eruption of the minifilament, the most significant process in the LOS magnetogram is the convergence and flux cancellation between two opposite magnetic polarities at the southwest end of the minifilament (Figure \ref{figmg}(a)--(e)). We track the evolution of these two polarities and plot their unsigned magnetic flux in the region enclosed by the blue box in Figure \ref{figmg}(f). The blue curve in Figure \ref{figmg}(f) shows three episodes of flux cancellation. The first cancellation lasts from 17:29 to 17:43 UT, the second one occurs from 17:52 to 17:59 UT, and the last one takes place from 18:05 to 18:14 UT. These three flux cancellations correspond well with the observations in H$\alpha$. The first flux cancellation occurs at the southwest end of the southern J-shaped arcade, while it is connecting with the external dark structure in the same location. The second flux cancellation corresponds to the onset of the coronal jets, and both jets originate from the region with flux cancellation. The last flux cancellation takes place at the southwest footpoint of the minifilament, where it connects with the second external dark structure and begins to erupt. Meanwhile, these two opposite polarities with flux cancellations continue to display a converging motion throughout the event, as visualized in Figure \ref{figmg}(g).

Figure \ref{figmg}(h) shows the photospheric flows during 17:36 - 17:44 UT when the southern J-shaped arcade vanishes and the sigmoidal structure forms. The photospheric transverse velocity map is derived using the differential affine velocity estimator (DAVE) code \citep{2006ApJ...646.1358S} by combining six successive magnetograms, in order to increase the signal-to-noise ratio. %We further calculate the magnetic helicity flux density caused by horizontal photospheric motion, by deriving vector potential $\boldsymbol{A_{p}}$ and calculating $-2(\boldsymbol{v\cdot A_{p}})B_{z}$ using the method described by \cite{2001ApJ...560L..95C}. 
In Figure \ref{figmg}(h), the evolved northern J-shaped arcade is traced in H$\alpha$ and superposed on the magnetogram, shown by cyan lines. Counterclockwise rotation is detected inside the region enclosed by the yellow circle. %, which injects negative helicity into the northern J-shaped branch, as shown by the helicity flux density contours. The total helicity flux in the FOV of Figure \ref{figmg}(h) is $-2.2\times 10^{36}~{\rm Mx^{2}~hr^{-1}}$. 
This rotational motion may inject twist into the northern J-shaped arcade, leading to the formation of the sigmoidal structure. Converging motion is seen in the region enclosed by the yellow box, which may be the precursor of the second flux cancellation and the onset of coronal jets.

\subsection{Interaction with the Overlying Large-scale Filament}
\subsubsection{The Failing of the minifilament Eruption}

\begin{figure*}[hbtp!]
\plotone{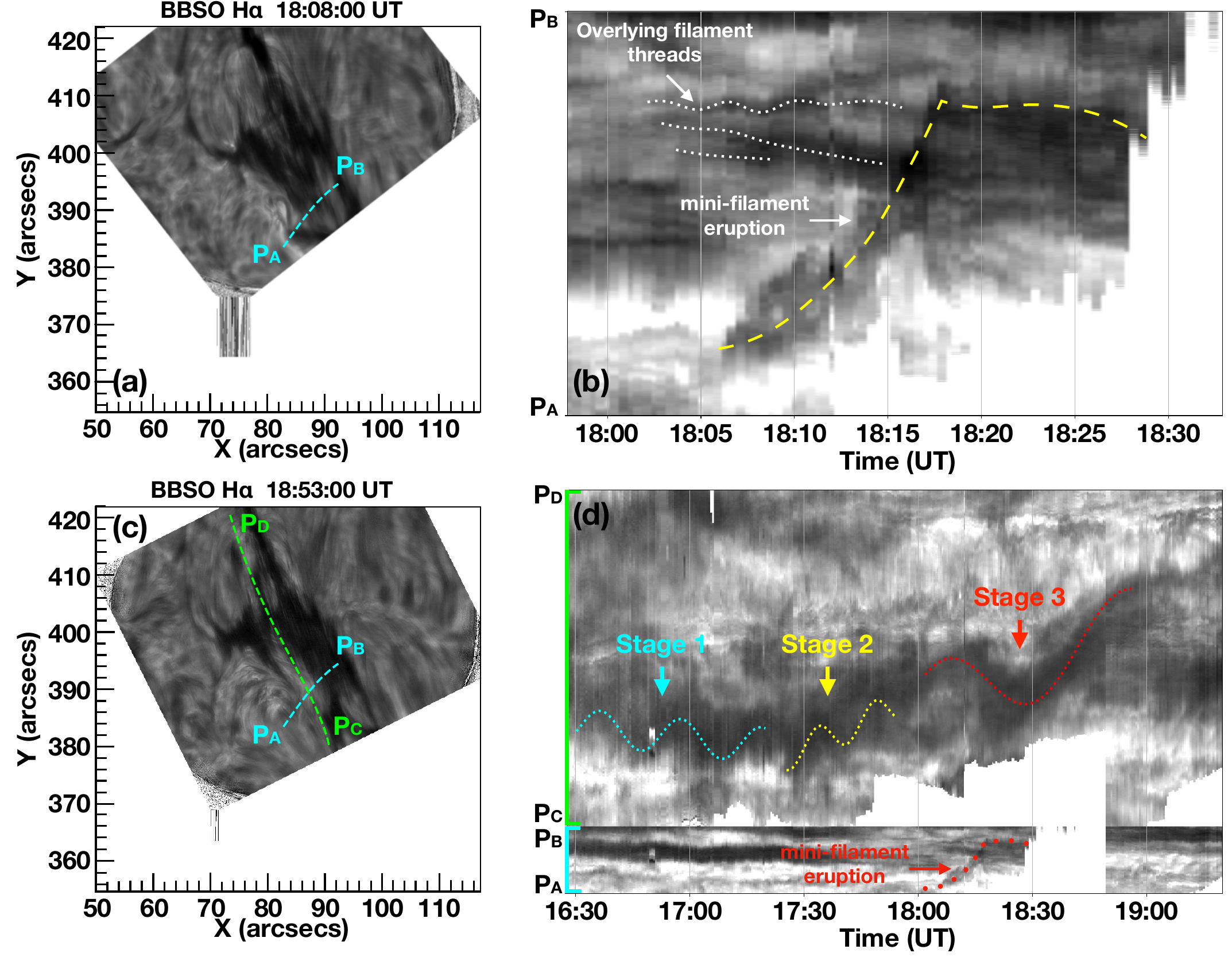}
\caption{The failing of the minifilament eruption (top row) and oscillation enhancement of the overlying large filament (bottom row). The locations of the slits are shown as dashed lines in the left panels. Corresponding distance-time plots along the slits shown in the left panel are presented in the right panels. The yellow dashed line in panel (b) outlines the motion of erupting minifilament, and the dark threads of the large overlying quiescent filament are marked by the white dotted lines. The lower and upper parts of panel (d) show the time-distance map along ${\rm P_{A}-P_{B}}$ and ${\rm P_{C}-P_{D}}$ in panel (c), respectively, and ${\rm P_{A}-P_{B}}$ is the same as the slit in (a). The cyan, yellow and red dotted lines in the upper part of panel (d) show three different stages of the filament oscillation, and the red dotted line in the lower part traces the minifilament eruption.}
\label{figos}
\end{figure*}

The failed eruption transforms the minifilament into an arcade that gradually fades away. The time-distance map in Figure \ref{figos} (b) clearly shows the sudden halt of the minifilament eruption. This may be partly due to the interaction and possible magnetic reconnection between the erupting minifilament and the overlying large quiescent filament. In addition, observations in AIA 171~\AA~and other EUV bands indicate that some of the erupting hot material flows out horizontally from the erupting site.

\subsubsection{Oscillation of the Overlying Large-scale Filament}

%In addition to eruptions, large filaments often exhibit a milder but more widespread behavior known as oscillation. There are generally two types of filament oscillations: longitudinal and transverse. Longitudinal oscillations occur when filament threads oscillate parallel to the main filament axis \citep{2003ApJ...584L.103J}. Transverse oscillations, on the other hand, occur when the filament oscillates in the plane perpendicular to the filament axis \citep{10.1093/mnras/91.2.239,2014ApJ...795..130S}. 

Large-scale filament oscillations can be triggered by coronal waves generated by distant eruptive events \citep{2013ApJ...773..166L,2015ApJ...814L..17S}, or nearby small-scale eruptions and subflares \citep{2012A&A...542A..52Z,2007A&A...471..295V,2003ApJ...584L.103J}. Longitudinal oscillations \citep{2003ApJ...584L.103J} occur due to the field-aligned component of gravity near magnetic dips, where the oscillation period is determined by the shape of the field line. The typical longitudinal oscillation period is about an hour, but it can be as short as 20 minutes when the curvature radius of the field lines near the magnetic dip is small \citep{2017ApJ...835...94O}.

In the case we are investigating, the interaction of the erupting minifilament with the overlying field not only leads to the failed eruption, but also causes disturbances of the overlying large-scale filament. In the time-distance map (Figure \ref{figos} (d)), we fit the motion of the overlying large-scale filament threads with a function $s=A{\rm sin}(\omega t+ b)+ct+d$. The selection of this function, which characterizes a longitudinal oscillation superimposed on an overall migrating motion, is motivated by the fact that the filament threads demonstrate a northward migration at a quasi-steady speed while exhibiting weakly damped oscillations, after 17:26 UT. By disregarding the damping of the oscillations, we aim to achieve a more accurate fit for both the oscillation amplitude and the migration speed. Before 17:26 UT, the threads are undergoing small amplitude oscillation with an amplitude of about 2.35 arcsecs, and the period is around 21 minutes. After 17:26 UT, when the southern branch of small-scale J-shaped arcades disappears, the threads start to migrate northwards at a speed of 2.78 ${\rm km~ s^{-1}}$, meanwhile they continue to oscillate mildly with a slightly smaller amplitude (2.0 arcsecs) and a shorter period (15 minutes). After 18:05 UT, when the minifilament erupts, the threads start to oscillate with a much larger amplitude (5.12 arcsecs) and a much longer period (47 minutes), while migrating northward with a speed of 2.38 ${\rm km~ s^{-1}}$. It should be noted that after the minifilament eruption, the oscillation is only observed for about one period, so the fitted migration speed, oscillation period and amplitude may have large uncertainties, but the amplification of oscillation is indeed detected. In conclusion, a northward migration is detected after the disappearance of the southern J-shaped arcade, and an amplified longitudinal oscillation is seen after the start of the eruption.

\section{Magnetic Field Modeling}
\label{sect:model}

\subsection{Flux Rope Insertion Method}

In order to understand the formation and eruption process of the minifilament, we construct magnetic field models using the flux rope insertion method developed by \cite{2004ApJ...612..519V}. This method has been successfully applied to model the source regions of large-scale events such as active region filaments \citep{2008ApJ...672.1209B, 2009ApJ...691..105S, 2009ApJ...704..341S, 2011ApJ...734...53S, 2018ApJ...855...77S}, quiescent filaments \citep{2012ApJ...757..168S, 2015ApJ...807..144S}, a double-decker filament \citep{2021ApJ...923..142C}, sigmoids \citep{2009ApJ...703.1766S, 2012ApJ...750...15S}, an erupting pseudostreamer \citep{2021cosp...43E1768K}, a blowout jet \citep{2022ApJ...938..150F} and so on. The smallest structure we have modeled so far is a mini sigmoid with $\sim$20$^{\prime\prime}$ in length \citep{2019ApJ...887..130H}.

A brief description of the method is below. We first compute the potential field from a high-resolution (HIRES) magnetogram embedded in a global synoptic map. Then a thin flux bundle representing the axial flux of the flux rope ($\Phi_{\rm axi}$) is inserted to the cavity created above the selected filament path. Circular loops representing the poloidal flux of the flux rope ($F_{\rm pol}$) are then added around the flux bundle. At last, the field is relaxed through magnetofrictional relaxation \citep{1986ApJ...309..383Y}. A series of models with different inserted magnetic fluxes are constructed, and the best-fit model is identified after comparison with observations, for details please refer to \cite{2009ApJ...691..105S, 2012ApJ...757..168S}.

%As the method starts with a potential field containing an inserted flux rope, the regions far from the inserted flux may maintain highly potential after the magneto-frictional relaxation, which may not reliably represent the real coronal magnetic field. But for the core structures near the inserted flux, the resultant magnetic configuration significantly deviates from the potential field. 
The magneto-frictional relaxation simplifies the MHD equations by neglecting the thermal pressure and gravity, and adds a frictional dissipation term $\boldsymbol{D(v)}=-\nu\boldsymbol{v}$ to the momentum equation. Therefore, the velocities during the relaxation are not real, and the timescale has no physical meaning. However, the spatial structures and magnetic topology after the relaxation are realistic, which is helpful for unveiling the magnetic configurations behind the observations. This method also fails to describe MHD waves or generate slender current sheets, but they are not needed in this study. The flux rope insertion method only requires LOS magnetograms, therefore it is suitable for modeling weak magnetic field in the quiet region, as in this case. In conclusion, the flux rope insertion method is applicable for modeling this minifilament.

\subsection{Comparison with the Formation of the Minifilament}

\begin{figure*}[hbtp!]
\plotone{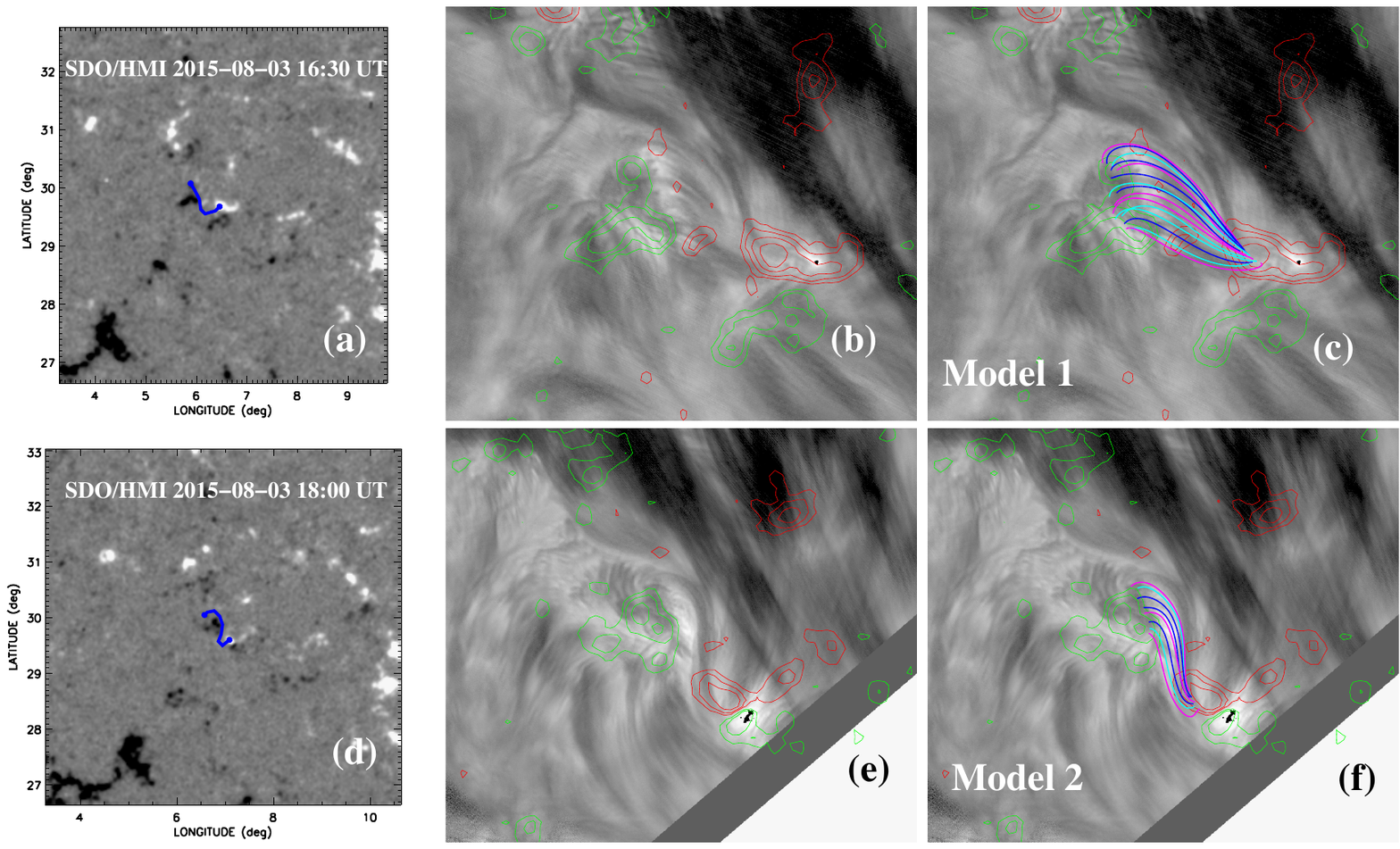}
\caption{Magnetic modeling of the minifilament at 16:30 UT (top row) and 18:00 UT (bottom row) before the eruption. Left column: Longitude–latitude maps of the radial component of photospheric magnetic field in the HIRES region of the model. The blue lines ended with two circles show the paths of the inserted flux bundles. Middle and Right columns: The corresponding H$\alpha$ images with a much smaller FOV are shown in the background. The color curves on the right column show the modeled core field structure corresponding to the observed minifilament. The red and green contours refer to the corresponding positive and negative magnetic fields observed by HMI.}
\label{fig9}
\end{figure*}

We have constructed a series of magnetic field models using the LOS photospheric magnetograms taken by SDO/HMI at 16:30 UT and 18:00 UT. The computation domains of the HIRES regions span about 7.3$^\circ$ in longitude and 6.4$^\circ$ in latitude as shown in the left row of Figure \ref{fig9}. The spatial resolution of the HIRES region and the global map in the low corona is 0.0005 $R_{\sun}$ and 1$^\circ$, respectively. Both regions extend from the solar surface up to a source surface of 1.7 $R_{\sun}$. For both the 16:30 UT and 18:00 UT best-fit models, the axial flux $\Phi_{\rm axi}$ of the inserted flux rope is $3\times 10^{19}$ Mx. The poloidal flux per unit length along the flux rope, $F_{\rm pol}$, is equal to $0~{\rm Mx~cm^{-1}}$ in the best-fit models, which means no poloidal flux is initially inserted in our model.

A comparison of the images in the middle and left rows of Figure \ref{fig9} shows that the selected field lines from the best-fit models after 30000-iteration relaxation match the observed dark filament threads well. The magnetic free energy in the model increases from $8.7\times 10^{27}$ erg in Model 1 at 16:30 UT to $9.6\times 10^{27}$ erg in Model 2 at 18:00 UT, while the corresponding potential field energy decreases from $1.473\times 10^{30}$ erg to $1.446\times 10^{30}$ erg, likely due to flux cancellation.

\subsection{Comparison with the Pre-eruption Coronal Jets}

\begin{figure*}[hbtp!]
\plotone{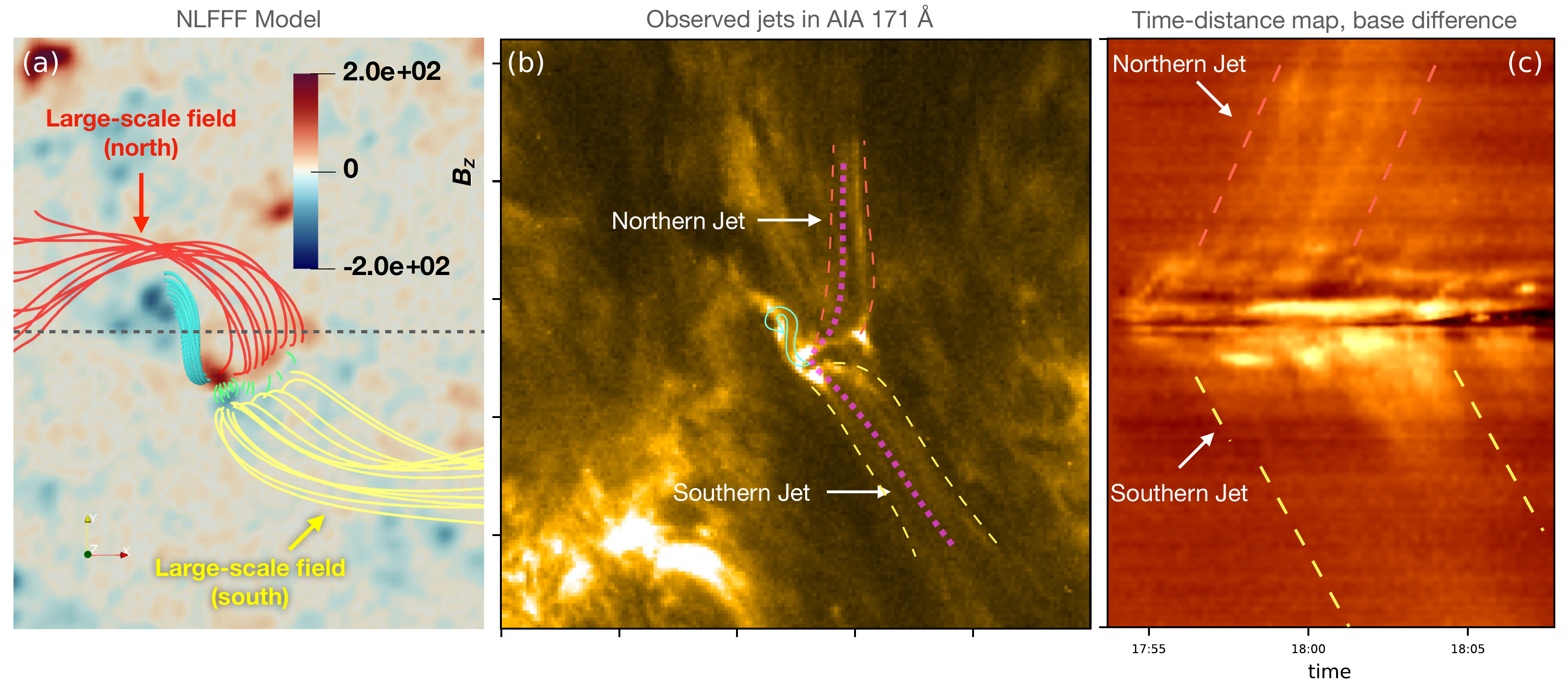}
\caption{(a) Magnetic field lines of the minifilament (cyan) and nearby larger-scale field lines (red and yellow) in the best-fit non-potential model at 18:00 UT. The red and blue patches refer to the positive and negative magnetic polarities observed by HMI. (b) EUV coronal jets observed in 171~\AA~by AIA. The red and yellow dashed lines mark the paths of two coronal jets, and the cyan line marks the H$\alpha$ minifilament. (c) Time-distance map of the twin coronal jets, along the magenta dotted slit in panel (b). The red and yellow dashed lines outline the region of each coronal jet.}
\label{figjtc}
\end{figure*}

The coronal magnetic field model at 18:00 UT provides important insights into the mechanism of the coronal jets observed before the minifilament eruption. In Figure \ref{figjtc}(a), we plot selected field lines corresponding to the observed minifilament and neighboring fields to compare them spatially with the observed coronal jets. The AIA 171~\AA~image and the time-distance plot of the base-difference image in Figures \ref{figjtc}(b) and \ref{figjtc}(c) show that the two coronal jets (marked with dashed yellow and orange curves) originated from the south end of the minifilament flow along opposite directions. A comparison of panels (a) and (b) of Figure \ref{figjtc} shows that two sets of large-scale model field lines (orange and yellow curves) match the two observed coronal jets relatively well in both direction and location. These field lines are rooted in closely-located opposite polarities at the southwest end of the minifilament. However, the alignment between the observed jets and magnetic field lines in the best-fit model is not perfect, which is likely due to the different overlying fields above the minifilament. In the observations, there is a large scale quiescent filament along the paths of coronal jets indicating existence of non-potential fields, while the large scale overlying fields in the model are potential. However, this discrepancy does not affect our understanding of the observations.

A combination of the observed magnetic flux evolution and magnetic field modeling suggests that the magnetic reconnection between the two sets of large-scale magnetic field lines corresponding to the second flux cancellation may trigger the EUV coronal jets. This mechanism is similar to those bi-directional coronal jets produced by magnetic reconnection suggested by \cite{1997Natur.386..811I,2020RAA....20..138N}, etc. However, it differs from those of standard jets or blow-out jets. The standard jet model requires the emerging field to reconnect with the pre-existing ambient field \citep{1992PASJ...44L.173S}, whereas in this event, no flux emerging is detected. While in the blow-out jet model, the sigmoidal core field should erupt together with the jets \citep{2010ApJ...720..757M}. However, in this event, the minifilament eruption occurs about ten minutes after, when the coronal jets almost disappear. Due to the same reason, the theory of minifilament eruption producing coronal jets \citep{2015Natur.523..437S} also cannot explain this event.

\subsection{Eruption Mechanisms}

After the coronal jets, the minifilament erupts. Many mechanisms for triggering filament eruptions have been proposed. In ideal MHD, kink instability \citep{2010A&A...516A..49T} and torus instability \citep{2006PhRvL..96y5002K} are the main triggering mechanisms. Under resistive conditions, when magnetic reconnection occurs, theories such as tether-cutting \citep{2001ApJ...552..833M} and magnetic breakout \citep{1999ApJ...510..485A} models are commonly considered for triggering large-scale eruptions. To determine which mechanism is responsible for the minifilament eruption, we first calculate the twist number ($N$) and the decay index ($n$) based on the reconstructed coronal magnetic field models at 18:00 UT.

\begin{figure*}[hbtp!]
\plotone{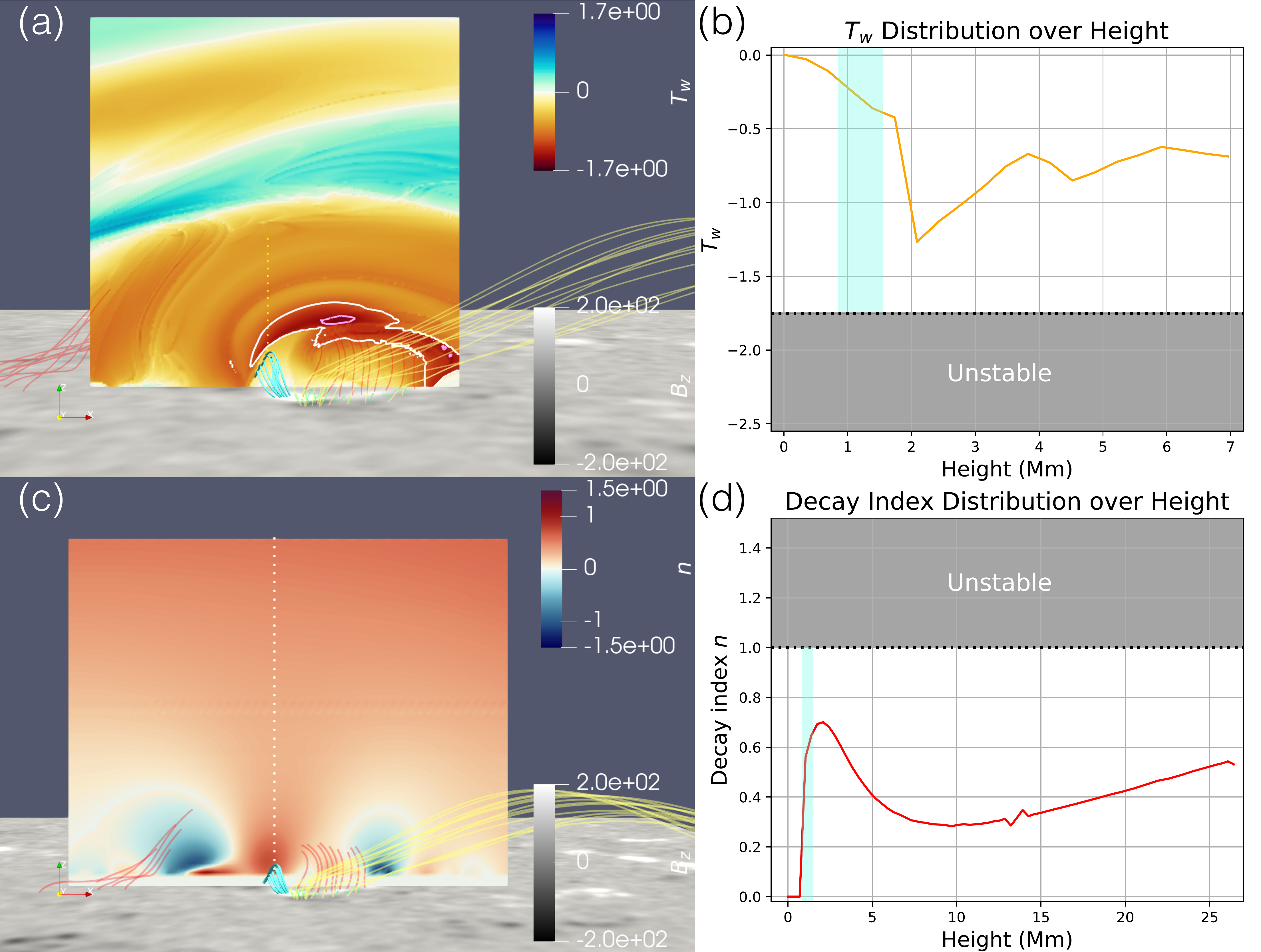}
\caption{(a) Twist ($T_{w}$) distribution in a cross section (marked by the black dashed line in Figure \ref{figjtc} (a)) of the modeled minifilament at 18:00 UT. Magnetic field lines are represented in the same way as Figure \ref{figjtc} (a). The white and pink contours corresponds to the $|T_{w}|$ value of 1.0 and 1.4, respectively. (b) $T_{w}$ distribution along the slit shown by the yellow dotted line in panel (a). Gray shaded area represents the kink-unstable region where $|T_{w}|>1.75$. Cyan shaded region represents the minifilament core field. (c) Similar to panel (a), but for distribution of decay index ($n$) calculated based on the potential field model at 18:00 UT. (d) Decay index distribution along the slit shown by the white dotted line in panel (c). Gray shaded area represents the unstable region where $n>1.0$ due to torus instability. Cyan shaded region is the same as panel (b).}
\label{figins}
\end{figure*}

The twist number is a parameter used to quantify the number of turns of a magnetic flux rope around its axis \citep{2019ApJ...884...73D}. We adopt the equation $T_{w}={\rm \int}(\boldsymbol{\nabla\times B})\cdot \boldsymbol{B}/(4\pi B^{2}){\rm d} l$ \citep{2006JPhA...39.8321B} to estimate the twist number by integrating along the field line and taking into account the length of the field line ($l$). $T_{w}$ is an approximation of the twist number in the vicinity of the flux rope axis \citep[][Appendix C]{2016ApJ...818..148L}. Figure \ref{figins}(a) shows a cross-section of the $T_{w}$ distribution in the best-fit model at 18:00 UT, and the sigmoidal core field supporting the minifilament is displayed in cyan lines. Figure \ref{figins}(b) shows the distribution of $T_{w}$ along the yellow dotted line in Figure \ref{figins}(a), and the location of the sigmoidal core field is marked by the cyan shading. We find that $|T_{w}|$ is below the threshold for triggering kink instability ($\Phi_{cr}=3.5\pi$ or $|T_{w}|=1.75$, according to \cite{2003ApJ...589L.105F,2004ApJ...609.1123F,2004A&A...413L..27T}). Therefore, kink instability unlikely plays a role in triggering the minifilament eruption. 

Under different physical conditions, different critical twist numbers are suggested for triggering kink instability. Here we justify our choice of $|T_{w}|=1.75$, or $\Phi_{cr}=3.5\pi$ as the critical value. \cite{1996ApJ...469..954L} studied the stability of a twisted horizontal flux tube under the photosphere. The flux tube is confined with external thermal pressure, and the magnetic field intensity at the outer boundary of the flux tube is set to zero. This physical condition is different from the situation in our study, where the minifilament, though small-scale, is over 1 Mm above the photosphere. It is a low plasma-$\beta$ environment where the magnetic pressure, rather than the thermal pressure, dominates the plasma dynamics. Therefore, the threshold in \cite{1996ApJ...469..954L} could not be used in this physically-different situation. \cite{1981GApFD..17..297H} perform a stability analysis of a line-tied, uniformly twisted and cylindrical flux tube, and find the critical value of about $|T_{w}|=1.25$ ($\Phi_{cr}=2.49\pi$) for triggering kink instability. However, the flux tube in that study is simplified to 2.5D and is straight rather than arched, which is different from the morphology of real flux tubes in the corona. To better simulate the development of kink instability of a realistic, three-dimensional and arched flux tube, \cite{2004ApJ...609.1123F} conducted a numerical simulation and find the critical twist is $\sim 1.76$ turns ($\Phi_{cr} \sim 3.5 \pi$) for triggering instability. The numerical analysis by \cite{2004A&A...413L..27T} also finds that $\Phi_{cr} \sim 3.5 \pi$ (twist $\sim 1.75$ turns) is the threshold for the kink instability of an arched flux rope in the corona. These realistic simulations corroborate each other, confirming that the threshold for an arched coronal flux tube is twist $\sim 1.75$ turns, or $\Phi_{cr} \sim 3.5 \pi$. Therefore, we believe that our choice of $|T_{w}|=1.75$ for the threshold of kink instability is reasonable.

The decay index ($n$) reflects the rate at which the external transverse field decreases with increasing height. A higher $n$ indicates a faster decrease in the downwards restricting force, making eruptions easier to occur. The decay index can be obtained through the following equation: $n=-{\rm \partial ln}B_{ex}/{\rm \partial ln}h$, in which $B_{ex}$ refers to the transverse flux density of the external field, and $h$ represents the distance to the solar surface. Figure \ref{figins}(c) shows a cross-section of the decay index distribution calculated based on the potential field model at 18:00 UT, and the sigmoidal core field is traced by cyan lines. In Figure \ref{figins}(d), we present the distribution of decay index $n$ {\tiny }along the yellow dotted line in Figure \ref{figins}(c). We find that the decay index in the domain and the vicinity of the sigmoidal core field is much smaller than the critical decay index ($n_{cr}\in [1.0,2.0]$) provided by theoretical calculations and numerical simulations \citep{2006PhRvL..96y5002K,2007ApJ...668.1232F,2010ApJ...708..314A,2010ApJ...718.1388D} for triggering torus instability. Therefore, torus instability also unlikely plays a role in the initiation of this event.

\begin{figure*}[hbtp!]
\plotone{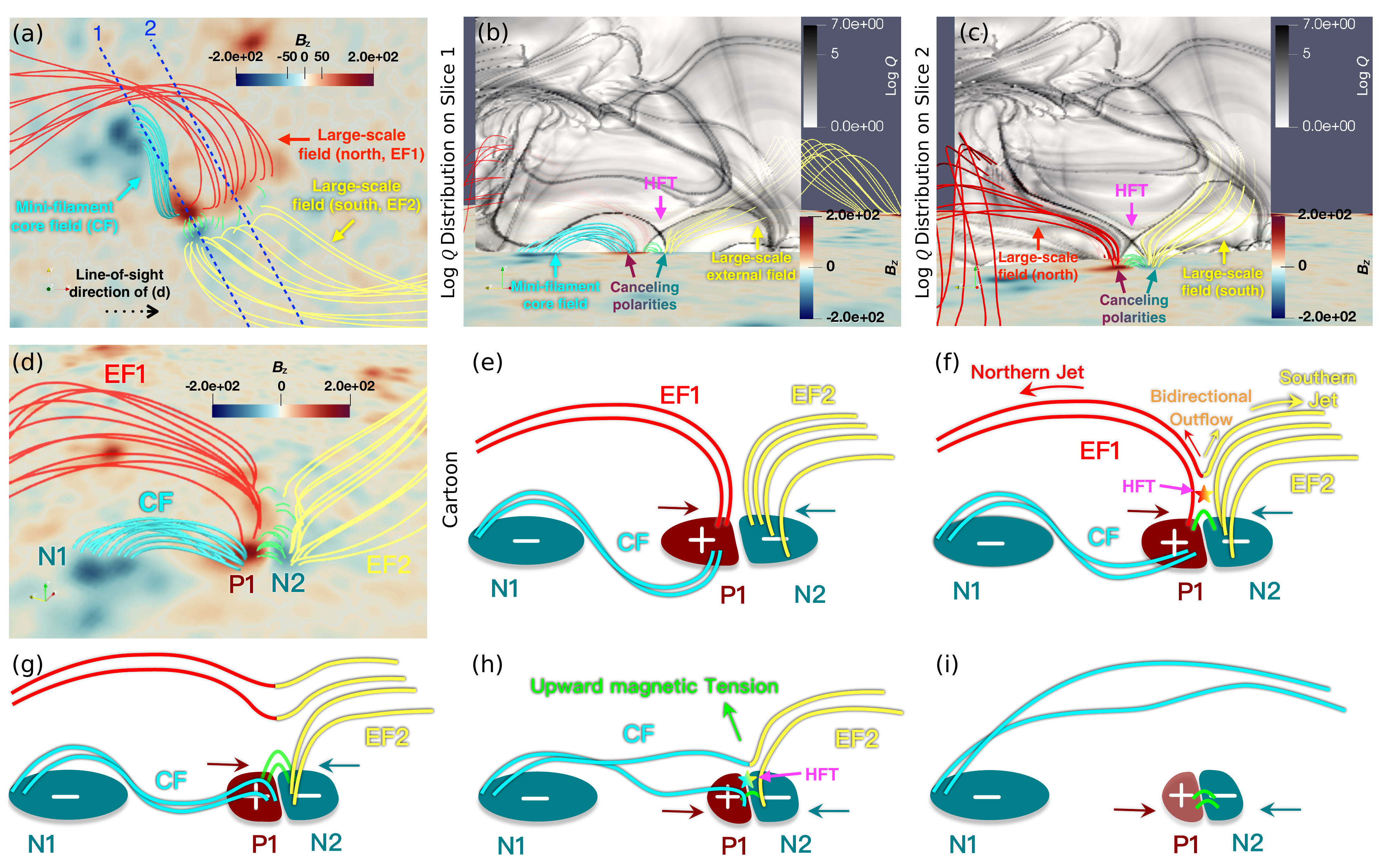}
\caption{Magnetic field modeling and cartoon of the minifilament eruption. (a) Top view of the magnetic field modeling result. (b)-(c) Log $Q$ maps of the best-fit model on the vertical slices marked by blue dashed lines 1 and 2 in panel (a), respectively. Key features are indicated with arrows and text. (d) Similar to panel (a), but for a side view of the magnetic modeling result. The red and blue patches in panels (a)-(d) refer to the positive and negative magnetic polarities observed by HMI. The modeled magnetic field lines in panels (a)-(d) are shown in the same way as those in Figure \ref{figjtc} (a).  (e)--(i) Cartoon for the minifilament eruption based on the modeling results and observation. The cyan, red, yellow, green lines represent the sigmoidal core field (CF), the northern external field (EF1), the southern external field (EF2) and the magnetic field below the HFT, respectively. The colored areas in the bottom of each panel refer to the photospheric magnetic polarities, and the arrows with the same colors represents their motions. The star signs mark the HFTs where magnetic reconnections take place. The red and yellow arrows along the field lines in panel (f) represent the reconnection outflows, and the green arrow in panel (h) indicates the direction of the magnetic tension force that lifts up the minifilament.}
\label{figct}
\end{figure*}

%After excluding the ideal MHD instabilities, we turn to magnetic reconnection. 
The third episode of flux cancellation at the southwest polarities (18:05 - 18:14 UT), the previous disappearance of the southern J-shaped arcade, the onset of two coronal jets, the connection with the external dark structure at the minifilament's southwest end, and the gradual eruption of minifilament from south to north, all suggest that magnetic reconnection occurs around the filament's southwest end. To examine this, we calculate the squashing factor $Q$ based on the reconstructed magnetic field. The places with a large squashing factor are quasi-separatrix layers (QSLs), where the linkage of neighboring magnetic field lines changes dramatically, thus providing advantageous conditions for current sheet development. Figures \ref{figct}(b)-(c) show the distribution of ${\rm log}_{10}Q$ in two different vertical slices marked by dashed lines in \ref{figct}(a). On both slices, we can see two QSLs crossing each other, forming a hyperbolic flux tube (HFT) above the cancelling southwest polarities, where magnetic reconnection prefers to take place. The two sets of large-scale field lines represented by the red and yellow lines are located on the two sides of the HFT. The sigmoidal core field represented by the cyan lines are located within the same domain as the red field lines.

To help understand the trigger and eruption mechanism of the coronal jets and minifilament eruption, we create a series of illustrative cartoon diagrams. Figure \ref{figct}(d) shows results from the best-fit magnetic field model at 18:00 UT, whose magnetic configuration is sketched in Figure \ref{figct}(e). On the photosphere, there are three  polarities namely, N1 (negative, northeast), P1 (positive, southwest) and N2 (negative, southwest). The two adjacent opposite polarities P1 and N2 are being squeezed together by converging flows and cancels with each other. In the corona, there are three sets of field lines, i.e., the small-scale sigmoidal field lines CF supporting the minifilament (cyan), and the two large scale external field lines EF1 (red) and EF2 (yellow) connecting to polarities P1 and N2, respectively. As the converging motion of P1 and N2 continues, EF1 and EF2 come close to each other and reconnect, producing bidirectional reconnection outflow that generates twin coronal jets (Figure \ref{figct}(f)), as observed in EUV.  Afterwards, the converging flow continues (Figure \ref{figct}(g)), and the sigmoidal field lines CF and EF2 are driven to reconnect with each other and form a bunch of large-scale field lines. The newly reconnected part between the two former field lines creates a deep `valley,' where the magnetic tension force is strong and upward (Figure \ref{figct}(h)). Therefore, the newly reconnected large-scale fields (cyan) carrying the minifilament materials erupt, until they are stopped by the interaction with the overlying large-scale quiescent filament (Figure \ref{figct}(i)). 

\section{Discussions} 
\label{sect:disc}

Synthesizing all the information provided by multi-wavelength observations and magnetic reconstruction, we give an outline of the entire event as follows. A group of sheared magnetic arcades lies between an isolated negative magnetic polarity and two closely located opposite polarities under a large quiescent filament. The initial structure in H$\alpha$ appears as a group of dark arched threads, which then evolves into two J-shaped arcades. The converging motion between the closely located opposite polarities leads to three episodes of flux cancellations. Associated with the first flux cancellation, one of the J-shaped arcades reconnects with the external large-scale field, then rises and becomes invisible. After that, the isolated negative polarity undergoes a rotational motion, leading to the transformation of the other J-shaped arcade into a sigmoidal structure. Next, the second episode of flux cancellation indicates the reconnection between two large-scale external fields. The hot plasma produced by the reconnection flows out along these large-scale field lines, forming a pair of bi-directional coronal jets. Following that, due to the third episode of flux cancellation, the sigmoidal minifilament reconnects with the external large scale magnetic fields, then gradually erupts from one end to the other. The erupting minifilament interacts with the overlying large filament and leads to the enhancement of the large filament's oscillations, and the eruption then fails partly because it hasn't reached the threshold height of torus instability.

The formation of this sigmoidal minifilament is a bit different from the large-scale ones. In this event, before the formation of the sigmoidal structure, external magnetic reconnection first destroys one of the two J-shaped arcades, and the other J-shaped arcade evolves into a sigmoidal structure later, likely due to the rotation of its footpoint. However, large-scale sigmoidal structures are usually formed due to tether-cutting reconnection between two J-shaped arcades  \citep{2001ApJ...552..833M,2009ApJ...698L..27T,2010ApJ...725L..84L,2011A&A...526A...2G,2014ApJ...789...93C}. This may suggest that the formation mechanisms of some small-scale sigmoidal structures are different from that of larger-scale ones. Moreover, the destruction of one of the J-shaped arcades in this event suggests that external field may have a greater influence on solar miniature structures. This is quite reasonable since the amount of magnetic flux of small-scale structures is lower than large-scale ones, and more comparable to the flux of the ambient external field.

The eruption mechanism of this sigmoidal minifilament is similar to that of a large-scale sigmoid erupted on 2012 July 12. \citet{2022ApJ...940...62L} has performed a data-constrained MHD simulation of this large-scale sigmoid, and found that the magnetic reconnection between the sigmoid and external fields at the external null point leads to the motion of the sigmoid's right footpoint at the eruption onset. The similarity of these two events with a scale difference of about 30 times further supports that solar eruptions can be explained by similar mechanisms, regardless of their scales. 

The external magnetic reconnection in the current event results in the liftoff of one footpoint of the minifilament, while the other footpoint remains, making it an asymmetric filament eruption. According to a statistical study of filament eruptions by \cite{2015SoPh..290.1703M}, more than one-third of large-scale filament eruptions are asymmetric ones. Large-scale filaments tend to have a long lifetime, large spatial length and diverse activities, complicating the investigation of their eruption mechanisms. And as \cite{2018ApJ...859....3M} points out, the study on minifilaments can cover the entire evolution process, providing knowledge about the pre-eruption and eruption mechanisms of their large-scale analogs. In our study, the mechanism of this asymmetric minifilament eruption provides an alternative for the initiation of large-scale asymmetric filament eruptions: the reconnection between the filament core field and the external field.

\section{Summary}
\label{sect:summ}
In this paper, we present high-resolution observations of the formation and failed eruption of a minifilament located below a large quiescent filament on 2015 August 3. Multi-wavelength observations reveal several accompanying activities, such as three episodes of flux cancellation between two opposite polarities, bi-directional coronal jets, and oscillation enhancement of the large quiescent filament due to the minifilament eruption. Two non-potential magnetic field models are constructed at two time instants before the eruption, using the flux rope insertion method. The two best-fit models reveal the 3D coronal magnetic fields at the beginning of the BBSO/GST observation and before the minifilament eruption. This minifilament is the smallest structure that we have successfully modeled using the flux rope insertion method so far, and the performance shows the potentiality and reliability of this method. In the pre-eruption model, an HFT is identified between the minifilament core field and the external larger-scale field. This configuration is situated above the cancelling magnetic polarities, indicating the occurrence of external magnetic reconnection. In addition, both the twist number of the minifilament core field and the decay index near the minifilament are below the threshold required for the onset of ideal kink and torus instabilities.\\

The observation and modeling of this small-scale minifilament provides us with insights about the mechanisms of solar small-scale activities. The major findings are as follows:\\

The eruption of a minifilament can be triggered by the external reconnection between the filament core field and external field. The high-resolution observation by GST, together with NLFFF coronal magnetic field modeling, presents strong evidence for the external magnetic reconnection between the core field of the minifilament and the larger-scale external field loops before the eruption. Therefore, after excluding the occurrence of ideal MHD instabilities, the trigger of the eruption is suggested to be this reconnection between the filament and external magnetic loops. This scenario is different from the break-out model \citep{1999ApJ...510..485A}, in which the reconnection firstly occurs between the differently oriented overlying fields above the filament, and the filament-carrying field is not involved in the reconnection at the start of the eruption. In previous investigations of minifilaments, reconnection process between the erupting filament and the ambient field is found to occur after and as a result of the initiation of minifilament eruption \citep[$e.g.$][]{2015Natur.523..437S,2023A&A...673A..83L,2023ApJ...942...86Y}, while flux cancellation underneath the minifilament initiates the eruption \citep[$e.g.$][]{2014ApJ...783...11A,2016ApJ...832L...7P,2018ApJ...859....3M,2020ApJ...902....8C,2023ApJ...942...86Y}, just like large-scale filaments. However, the flux cancellation in the present event occurs between one footpoint of the minifilament and an external opposite polarity, not under the filament. Therefore, in this particular case, we show another possible scenario of minifilament eruption process: the minifilament first reconnects with an external magnetic loop, then rises and erupts, likely driven by the upward magnetic tension force of the post-reconnection field lines. This scenario also provides an alternative for large-scale asymmetric filament eruptions.\\%, as shown in the Discussions section.\\

A new relationship between minifilaments and coronal jets is presented. \cite{2015Natur.523..437S} suggests that coronal jets are caused by minifilament eruptions, which is supported by observations \citep[$e.g.$][]{2011ApJ...738L..20H,2014ApJ...783...11A,2016ApJ...830...60H,2022ApJ...927...79B}. Although the hot outflows of some jets start before the triggering of the minifilament eruption \citep[$e.g.$][]{2016ApJ...830...60H,2016ApJ...827...27Z}, they are caused by the break-out reconnection above the minifilament \citep{2016ApJ...827...27Z,2018ApJ...852...98W}, which accompanies the slow-rise phase of the minifilament. However, in this event, the bi-directional EUV coronal jets initiate and fade away before the eruption of the minifilament, meanwhile the minifilament stays stationary and is not disturbed. These jets are caused by the magnetic reconnection between two larger-scale external magnetic loops, driven by photospheric flux convergence and cancellation of the same opposite polarities that drive the subsequent eruption of the minifilament. Therefore, in this event, the EUV jets and minifilament eruptions are two relatively independent results of the same cause - flux cancellations. Different episodes of flux cancellation between the same two polarities drive the magnetic reconnection in which different magnetic structures are involved, resulting in minifilament eruption (when minifilament core field is involved) or EUV jets (when large-scale magnetic loops are merged). The EUV jets are not produced by the eruption of the minifilament, as the minifilament erupts about ten minutes after the initiation of EUV jets. On the contrary, the appearance of jets can be viewed as a precursor to the eruption of this minifilament, since it indicates flux cancellation and magnetic reconnection near the filament footpoint, which may later erode the minifilament core field, causing its eruption.\\

%Magnetic structures in the solar atmosphere are interrelated, meaning that changes in one region can impact nearby structures, such as sympathetic filament eruptions \citep{2017ApJ...844...70L}. Furthermore, disturbances at small scales may trigger nonlinear growth and eventually lead to large-scale solar eruptions \citep{2021ApJ...923...74D}. In the event we have investigated, the eruption of a minifilament has enhanced the oscillation and migration of the overlying large-scale filament. Given that high-amplitude oscillations may lead to quiescent filament eruptions, the interaction between minifilament and large-scale quiescent filament highlights the potential for minifilament eruptions to trigger large-scale activities \citep{2014ApJ...790..100B,2020ApJ...898...34F,2021ApJ...923...74D}. Therefore, it is essential to conduct further high-resolution observations and simulations of solar miniature structures, since they can impact space weather by promoting large-scale activities. 

\textbf{Acknowledgements} This work is supported by the National Key R$\&$D Program of China 2021YFA1600502 (2021YFA1600500), 2022YFF0503001 (2022YFF0503000), the NSFC (12173092,  11925302, 12273101, 11790302 (11790300)), and the Strategic Priority Research Program on Space Science, CAS, grant Nos. XDA15052200 and XDA15320301. Thanks for the sponsorship provided by the Natural Science Foundation of Xinjiang Uygur Autonomous Region for Outstanding youth and the High-level Flexible Talent Program of Xinjiang Uygur Autonomous Region. We gratefully acknowledge the use of data from the Goode Solar Telescope (GST) of the Big Bear Solar Observatory (BBSO). BBSO operation is supported by US NSF AGS-2309939 and AGS-1821294 grants. GST operation is partly supported by the Korea Astronomy and Space Science Institute and the Seoul National University. W. C. acknowledges support from US NSF grants - AST-2108235, AGS-2309939 and 1821294. Weilin Teng would like to express gratitude to Dr. Zhengxiang Li for providing caring guidance and spiritual support throughout his time at university. Additionally, Dr. Jianpeng Guo is acknowledged for introducing Weilin to the beauty and foundational concepts of solar physics, which ultimately led him to pursue research in this field. Special thanks are also extended to Zhengyuan Tian for engaging in beneficial discussions.

%continuous converging motion is detected
%Note that because of the shearing motion of these polarities, the map cannot show the whole picture of the three flux cancellations. During the last flux cancellation, the positive polarity moves away from the slit so it is not captured.
%\begin{figure*}[htbp!]
%\plotone{velocity.pdf}
% \caption{The motion of the magnetic polarities in two different time periods. Arrows show the velocity of every pixel, and the top left horizontal arrow shows the scale of 1 km s$^{-1}$. In the left panel, green means westward motion, while red stands for eastward. Cyan lines trace the H$\alpha$ arched threads. In the right panel, northward and southward motions are presented by yellow and cyan, respectively, while red $\&$ green contours show magnetized region with outward vorticity (counter-clockwise motion) and inward vorticity (clockwise motion), respectively. The two J-shaped strips in H$\alpha$ are traced by green lines.}
 %\label{figve}
%\end{figure*}

%The black dashed line in the middle show the plane position of Figure \ref{figins}

%from the direction marked by white arrow in (a)

\end{document}